\documentclass[twocolumn,showpacs,preprintnumbers,amsmath,amssymb,superscriptaddress,floatfix,nofootinbib]{revtex4}
\usepackage{graphicx}
\usepackage{epsfig}
\usepackage{epstopdf}
\usepackage{hyperref}
\usepackage{amsmath}
\usepackage{amsfonts}
\usepackage{amssymb}
\begin{document}
\title{Triangle singularity in $B^-\to K^-X(3872);X\to \pi^0\pi^+\pi^-$ and the X(3872) mass }
\date{\today}
\author{Raquel Molina} 
 \affiliation{Universidad Complutense de Madrid \& IPARCOS, Facultad de F\'isica, Departamento de F\'isica Te\'orica, Plaza Ciencias, 1, 28040 Madrid, Spain}
 
\author{Eulogio Oset}
  \affiliation{Departamento de F\'isica Te\'orica and IFIC, Centro Mixto Universidad de Valencia - CSIC, Institutos de Investigaci\'on de Paterna, Aptdo. 22085, 46071 Valencia, Spain}

\begin{abstract}
We evaluate the contribution to the $X(3872)$ width from a triangle mechanism in which the $X$ decays into $D^{*0}\bar{D}^0 -cc$, then the $D^{*0} (\bar{D}^{*0})$ decays into $D^0 \pi^0$ ($\bar{D}^0 \pi^0$) and the $D^0 \bar{D}^0$ fuse to produce $\pi^+ \pi^-$. This mechanism produces an asymmetric peak from a triangle singularity in the $\pi^+ \pi^-$ invariant mass with a shape very sensitive to the $X$ mass. We evaluate the branching ratios for a reaction where this effect can be seen in the $B^- \to K^- \pi^0 \pi^+ \pi^-$ reaction and show that the determination of the peak in the invariant mass distribution of $\pi^+ \pi^-$ is all that is needed to determine the $X$ mass. Given the present uncertainties in the $X$ mass, which do not allow to know whether the $D^{*0} \bar{D}^0$ state is bound or not, measurements like the one suggested here should be most welcome to clarify this issue.
\end{abstract}

\maketitle
\section{Introduction}
The X(3872) (now branded as $\chi_{c1}(3872)$ in the PDG~\cite{pdg}) was the first exotic state of the X, Y, Z series, observed in the $B\to KX(3872)$; $X(3872\to \pi^+\pi^-J/\psi$ \cite{belle} reaction and later on in different processes \cite{pdg}. The value of its mass, as listed in the PDG~\cite{pdg}, is
\begin{equation}
 M_X=3871.69\pm 0.17~\mathrm{MeV}\ ,\label{eq:xmass}
\end{equation}
which turns out to be compatible with the $D^{*0}\bar{D}^0$ threshold mass,
\begin{equation} m_{D^0}+m_{D^{*0}}=3871.68\pm 0.07~\mathrm{MeV}\ .\nonumber\end{equation}
Hence, it could be equally bound or unbound in the above channel. The amount of theoretical work devoted to this resonance is large and we divert the reader to the detailed discussion done in Ref.~\cite{slzhu}. The proximity to the $D^*\bar{D}^0$ threshold has led to the suggestion that this resonance could be a $D^*\bar{D}-cc$ state. Concretely, in Ref. \cite{wong} two independent bound states for $D^0\bar{D}^{*0}$ and $D^+D^{*-}$ were obtained at $3863.67~\mathrm{MeV}$ as $3871.77$ MeV respectively, but it was also suggested that these components could get mixed to give two $I=0,J=1$ states\footnote{Where $I$ and $J$ denote isospin and spin respectively.}, with the $I=0$ state corresponding to the X(3872). In Ref.~\cite{mehen} the X(3872) is associated to a $D^{*0}\bar{D}^0-cc$ state, while coupled channel calculations including both, the $D^{*0}\bar{D}^0-cc$ and $D^{*+}D^--cc$ channels, reproduce the X(3872) as an approximate $I=0$ combination of the neutral and charged channels for the wave function at short distances \cite{gamermann,juan,isodani}. One could understand the latter picture as follows: since the $X(3872)$ is bound by about $7~\mathrm{MeV}$ in the  charged $D^{*+}D^--cc$ component, this amount of binding energy contributes to make the system stable, being both components, neutral and charged, relevant for the whole system.

One method to improve the determination of the X(3872) mass was proposed in Ref.~\cite{guo} using a triangle singularity (TS) which appears in $D^{*0}\bar{D}^{*0}\to\gamma X(3872)$, where the $D^{*0}\bar{D}^{*0}$ would be produced by some source. A possible implementation of the idea is given in Ref.~\cite{braepem} with the $e^+e^-\to\gamma X(3872)$ reaction. The reason for this sensitivity of the reaction to the X(3872) mass is the sharp peak produced by the TS when the width of the intermediate state particle, $D^{*0}$ in this case, is small (of the order of $60~\mathrm{KeV}$).

\begin{figure}
 \centering
 \includegraphics[scale=0.7]{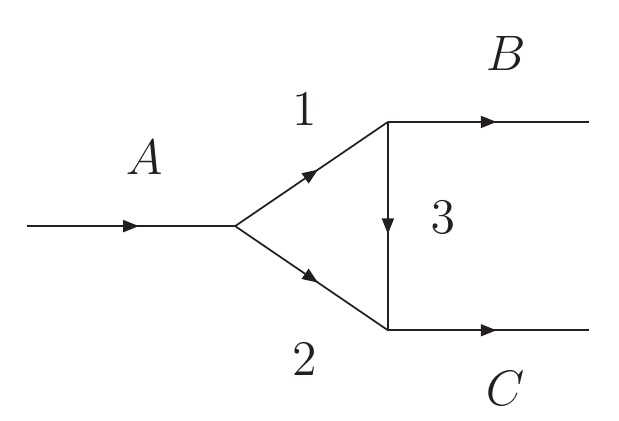}
 \caption{Feynman diagram for the reaction $A\to BC$ source of the triangle singularity (TS) phenomena.}
 \label{fig:figtri}
\end{figure}

Triangle singularities, popular in the sixties \cite{karplus,landau,booth,anisovich}, stem from reaction mechanisms where a particle $A$ decays into two particles $1+2$, then particle $1$ decays into particles $3$ and $B$, and latter $2+3$ fuse to produce a particle $C$. Altogether one has $A\to B+C$ with a triangle loop in the Feynman diagram with the particles $1,2,3$ (see Fig.~\ref{fig:figtri}). The appearance of the singularity demands that: $a)$ all particles $1,2,3$, are placed simultaneously on shell; $b)$ the latter are also collinear; and $c)$ the process can occur at the classical level (Coleman-Norton Theorem) \cite{coleman}. Formulated with a general framework using the Feynman parameterization of the amplitudes \cite{landau, qzhao}, the formalism can be done in a different way, which is technically easier and more transparent \cite{bayarguo}. In Ref.~\cite{bayarguo}, all the conditions for the TS are condensed in a single equation,
\begin{equation}
 q_\mathrm{on}=q_{a_-}\ ,
\end{equation}
where $q_\mathrm{on}$ is the momentum of particles $1,2$ in the $A$ rest frame, when they are placed on shell, and $q_{a_-}$, one of the two solutions for the momenta of particle $2$ when $2, 3$ are placed on shell to produce particle $C$, and $1$ and $B$ are parallel (see Ref.~\cite{bayarguo} for the expressions of $q_{\mathrm{on}}$ and $q_{a_-}$).

 At the beginning, there was a search for reactions which would show some enhancement due to a TS without much success \cite{booth,anisovich}. However, the situation has been reverted recently given the vast amount of experimental information gathered. One of the most relevant examples has been the interpretation in terms of a TS \cite{qzhao,mikha,acetidai} of the peak observed at COMPASS \cite{compass}, which originally was branded as a new resonance, the ``$a_1(1420)$''.

It was also relevant the explanation in terms of a TS \cite{wu,acetiliang,wuwu,acha} of the enhancement of the isospin forbidden decay of $\eta(1405)\to \pi^0f_0(980)$ \cite{beseta}. Another example is the explanation of an enhancement of the $\gamma p\to K\Lambda(1405)$ cross section around $\sqrt{s}=2110~\mathrm{MeV}$ \cite{moriya} solved in terms of a TS in \cite{wanguo}, or the interpretation of the $\pi N(1535)$ production channel \cite{vinisakai} in the $\gamma p\to p\pi^0\eta$ reaction \cite{gutz}.

Many examples of proposed TS to learn about the nature of some resonance, or to enhance dynamically suppressed production modes in different reactions, have been given in Refs.~\cite{guoliu,liangtri,daitri}. 

In the present work we go back to the work on the $X(3872)\to\pi^0\pi^+\pi^-$ reaction proposed in Ref.~\cite{achapion} and study in detail the $B^-\to K^-X(3872)$; $X(3872)\to\pi^0\pi^+\pi^-$ reaction showing that there is a TS for an invariant mass of $\pi^+\pi^-$ around $3729.7~\mathrm{MeV}$ with a peak quite sensitive to the mass of the $X(3872)$. The reaction proposed is similar to the one where the $X(3872)$ was originary found, $B\to KX(3872)$; $X\to J/\psi \pi^+\pi^-$, but the $J/\psi$ is now replaced by a $\pi^0$ and the rest is the same. The reaction has the novelty that because  there is a triangle singularity peak establishing a correspondence between the $X$ mass and the $M_{\pi^+\pi^-}$,  the $\pi^0$ does not have to be measured in principle. For practical purposes the measurement of the $\pi^0$ helps to reduce background, however, it is sufficient to know that a $\pi^0$ is produced without knowing with precision its energy and momentum.
\section{Formalism}
\subsection{Mechanism for the triangle singularity in the $B^-\to K^-X;X\to\pi^0\pi^+\pi^-$ reaction}
\begin{figure*}
 \centering
 \includegraphics[scale=0.7]{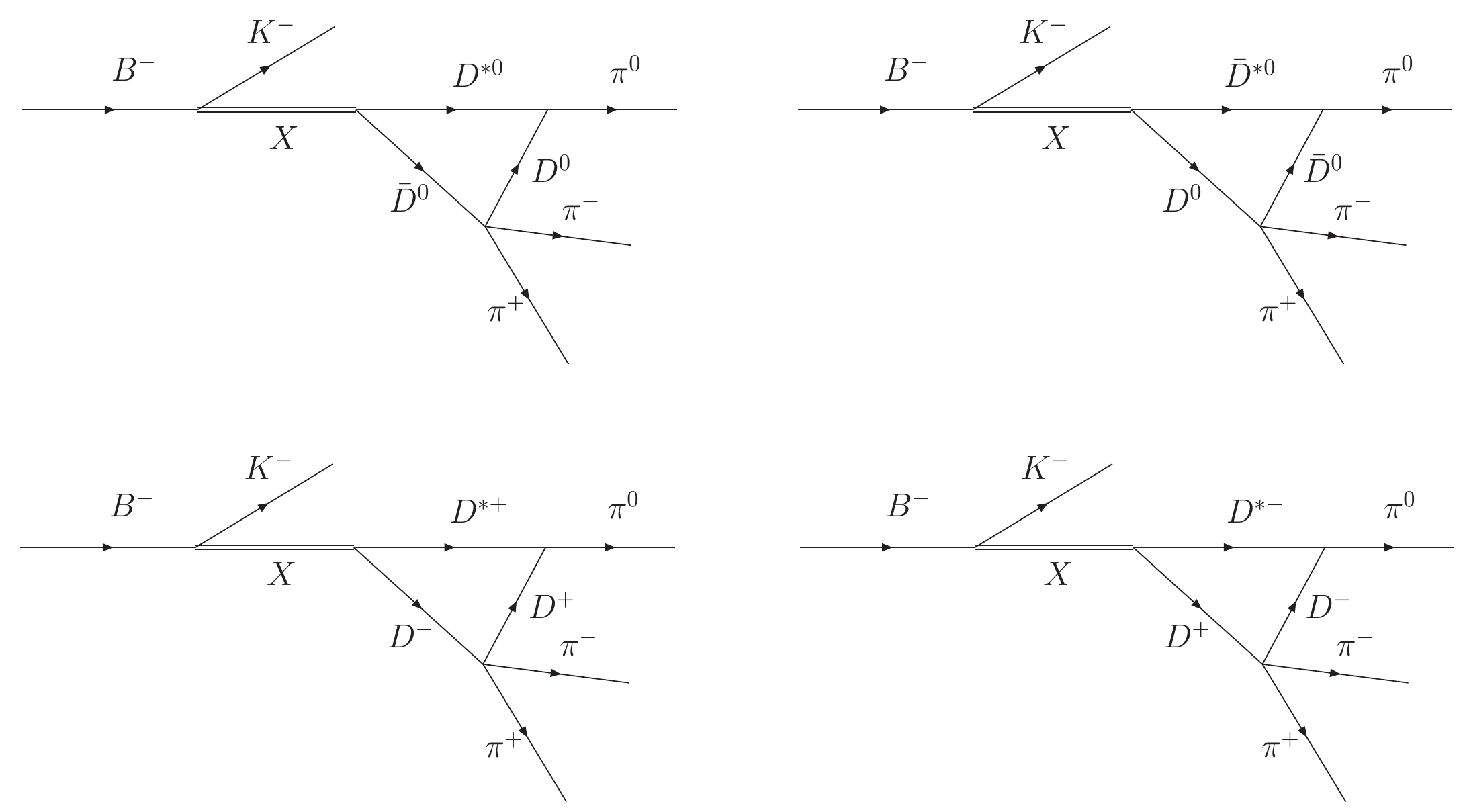}
 \caption{Mechanisms for $B^-\to K^-X;X\to \pi^0\pi^+\pi^-$ which develop a TS in the $\pi^+\pi^-$ invariant mass.}
 \label{fig:fig1}
\end{figure*}
The mechanism discussed above is depicted diagrammatically in Fig. \ref{fig:fig1}.
For the evaluation of the mass distribution of the $B^-\to K^-X;X\to\pi^0\pi^+\pi^-$ decay, several ingredients are needed:
\begin{itemize}
 \item[1)] the $B^-\to K^-X$ weak vertex,
 \item[2)] the coupling of the $X$ to the $D^*\bar{D}-cc$ state,
 \item[3)] the decay width for the $D^*\to D\pi$ process,
 \item[4)] the amplitude $D\bar{D}\to \pi^+\pi^-$, see Fig. \ref{fig:fig1}.
\end{itemize}
We proceed step by step to their evaluation.
\subsubsection{The $B^-\to K^-X$ vertex}
Since the $B^-$ and $K^-$ have both $J^P=0^-$, while $X$ has $J^{PC}=1^{++}$, $p$-wave is needed to compensate, because of the conservation of the angular momenta. Hence, the suited operator producing a scalar amplitude is 
\begin{equation}
 t_{B^-\to K^-X}={\mathcal C}\vec{\epsilon}_X\cdot \vec{p}_{K^-}\ ,
\end{equation}
where ${\mathcal C}$ is an unknown constant and has to be evaluated. There are several sources of experimental information. For example, Refs.~\cite{gobelle, pdg} give
\begin{equation}
 \mathcal{BR}(B^-\to K^-X;X\to D^0\bar{D}^0\pi^0)=(1.0\pm 0.4)\times 10^{-4}\ .
\end{equation}
Assuming that the $\mathcal{BR}$ of $X\to D^0\bar{D}^0\pi^0$ is about $50\%$, as deduced in Ref.~\cite{changzheng} we obtain the branching ratio
\begin{equation}
 \mathcal{BR}(B^-\to K^-X)\simeq (2.0\pm 0.8)\times 10^{-4}\ .\label{eq:br1}
\end{equation}
This number is in agreement with the one deduced in Ref.~\cite{changzheng} of $(1.9\pm 0.6)\times 10^{-4}$, and the one measured by BABAR, $(2.1\pm 0.6 \pm 0.3)\times 10^{-4}$ \cite{pdg}. We shall take Eq. (\ref{eq:br1}) for evaluations. 

The decay width of $B^-\to K^-X$ is then given by
\begin{equation}
 d\Gamma_{B^-}=\frac{1}{8\pi}\frac{1}{M^2_{B^-}}p_{K^-}\bar{\sum}\sum\vert t_{B^-,K^-X}\vert^2\label{eq:dgamb}
\end{equation}\vspace{0.2cm}
with $\bar{\sum}\sum \vert t_{B^-,K^-X}\vert^2=\mathcal{C}^2\vert\vec{p}_{K^-}\vert^2$, $p_{K^-}=\lambda^{1/2}(M^2_{B^-},m^2_{K^-},m^2_X)/2M_{B^-}$, and then,
\begin{equation}
 \frac{\mathcal{C}^2}{\Gamma_{B^-}}=\frac{8\pi M^2_{B^-}\mathcal{BR}(B^-\to K^-X)}{p^3_{K^-}}\label{eq:c}
\end{equation}
\vspace{0.2cm}
\subsubsection{The $X\to D^*\bar{D}-cc$ coupling}
We follow here the formalism of Refs.~\cite{gamermann,isodani}. The wave function at short distances is given there approximately by
\begin{equation}
 X\equiv\frac{1}{2}(D^{*+}D^-+D^{*0}\bar{D}^0-D^{*-}D^+-\bar{D}^{*0}D^0)\label{eq:xiso}
\end{equation}
which corresponds to $I^G(J^{PC})=0^+(1^{++})$ with the isospin phase convention, $(D^+,-D^0)$, $(\bar{D}^0,D^-)$, and the same for $D^*$\footnote{The $C-$ parity is acting over the charged mesons as $CD^+=D^-$, $CD^{*+}=-D^{*-}$}. In the diagrams of Fig. \ref{fig:fig1}, the diagrams for the TS after $B$ decays into $KX$ are depicted.  The $X$ coupling to all the components of Eq. (\ref{eq:xiso}) has to be evaluated. However, the $D^{*+}D^--cc$ components are bound by $7~\mathrm{MeV}$ for the $X$ mass, which means that the intermediate charged $D,D^*$ mesons in the loop can not be put on shell. This binding has to be compared to the $D^{*+}$ width of $(83.4\pm 1.8)~\mathrm{KeV}$~\cite{pdg}, hence we can say that they are very off shell and do not produce any appreciable contribution for the TS in the $X$ mass range (this conclusion is also found in Ref.~\cite{achapion}). Thus, the most important contribution comes from the $D^{*0}\bar{D}^0$ and $\bar{D}^{*0}D^0$ couplings, which just have opposite sign.

In Ref.~\cite{gamermann} the coupling of $X$ to the $D^*\bar{D}-cc$ state is evaluated, but the binding obtained is larger than the experimental one \cite{pdg}. A better estimate can be obtained using the Weinberg compositeness condition, see Refs.~\cite{weinberg,baru,danijuan}, which in the normalization used here is given by \cite{danijuan},
\begin{equation}
 g^2_X=\frac{16\pi s}{\mu}\sqrt{2\mu E_B}\label{eq:gx}
\end{equation}
being $s=M^2_X$, $\mu$ denotes the reduced mass of the $D^*,D$, and $E_B$ the binding energy of $X$ with respect to $D^{*0}\bar{D}^0$ system. With the value of $M_X$ given in Eq. (\ref{eq:xmass}) we compromise with the coupling, $g_X=2~\mathrm{GeV}$\footnote{This corresponds to binding energies of the $X$ around $20$ KeV. However, note that the position of the peak related to the TS is not altered by this value.}, and then,
\begin{eqnarray}
 g_{X,D^{*0}\bar{D}^0}=\frac{1}{2}g_X;\quad g_{X,\bar{D}^{*0}D^0}=-\frac{1}{2}g_X\ .\label{eq:gxv}
\end{eqnarray}
We should note that Weinberg's formula, Eq. (\ref{eq:gx}), holds for bound states. Our formalism can be used with unbound $D^{*0}\bar{D}^0$ components since we work in coupled channels and the $D^{*+}\bar{D}^--cc$ components are bound, stabilize the system and lead to a coupling of $X$ to the neutral components in the coupled channel approach. 

The full vertex function for the $X\to D^{*0}\bar{D}^0$ is then given by \cite{gamermann}
\begin{equation}
 t_{X,D^{*0}\bar{D}^0}=\frac{1}{2}g_X\vec{\epsilon}_X\cdot\vec{\epsilon}_{D^*}
\end{equation}
\subsubsection{The $D^*\to D\pi$ coupling}
We write for convenience
\begin{equation}
 t_{D^{*0},\pi^0D^0}=-\frac{\tilde{g}}{\sqrt{2}}(\vec{p}_{\pi^0}-\vec{p}_{D^{*0}})\cdot \vec{\epsilon}_{D^*}\ .\label{eq:tds}
\end{equation}
Then, taking the $D^{*+}$ decay width from the PDG \cite{pdg}, and using isospin symmetry together with the fact that the branching ratio for the $D^{*0}\to D^0\pi^0$ process is $64.7$\%, one obtains,
\begin{eqnarray}
 \tilde{g}=8.43;\quad \Gamma_{D^{*0}}=55.9~\mathrm{KeV} \ .
\end{eqnarray}
The value obtained here for $\Gamma_{D^{*0}}$ is similar to the one reported in \cite{bradstar} of $\Gamma_{D^*}\simeq 60~\mathrm{KeV}$. The coupling of $\bar{D}^{*0},\pi\bar{D}^0$ can be obtained from Eq. (\ref{eq:tds}) changing the direction of the lines in the corresponding Feynman diagram, and hence, one obtains a relative minus sign. This means that, together with Eq. (\ref{eq:gxv}), the first two diagrams of Fig. \ref{fig:fig1} give the same contribution.
\subsubsection{The $D\bar{D}\to\pi^+\pi^-$ amplitude}
For this amplitude we follow the approach of Refs. \cite{gamerscalar,gamerzou}, where, working in coupled channels, one finds a $D\bar{D}$ bound state corresponding to a pole at $\sqrt{s}_0=(3722-i 18)~\mathrm{MeV}$.

We shall find the TS at $M_{\pi^+\pi^-}$ very close to this energy. Thus we can use the pole expression for the amplitude,
\begin{equation}
 t_{ij}=\frac{g_ig_j}{s-s_0+i\sqrt{s_0}\, \Gamma}\ ,\label{eq:tx1}
\end{equation}
with $s_0=(3722~\mathrm{MeV})^2$, $\Gamma=36~\mathrm{MeV}$, $g_{D^+D^-}=(5962+i1695)~\mathrm{MeV}$, $g_{D^0\bar{D}^0}=(5962+i1695)~\mathrm{MeV}$, $g_{\pi^+\pi^-}=(9+i83)~\mathrm{MeV}$.

\begin{figure}
 \centering
 \includegraphics[scale=0.7]{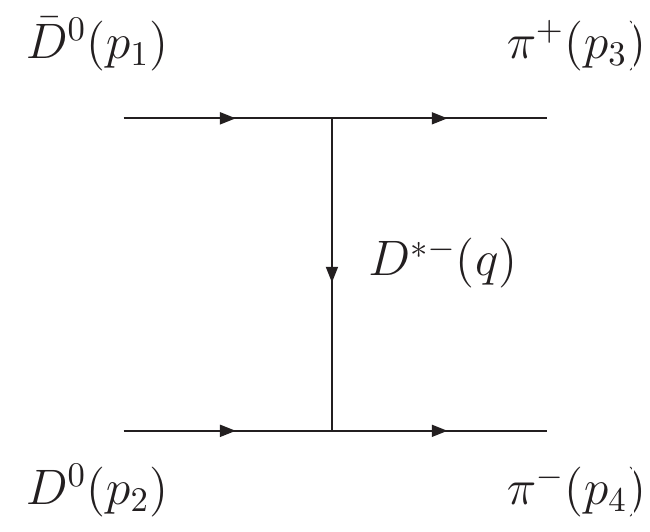}
 \caption{Feynman diagram for the perturbative amplitude $D^0\bar{D}^0\to \pi^+\pi^-$. In parenthesis the momenta of the particles.}
 \label{fig:fig2}
\end{figure}

We should note that the $D^0\bar{D}^0\to\pi^+\pi^-$ amplitude, in spite of being tied to a bound state, is much smaller than the perturbative amplitude given by the diagram in Fig. \ref{fig:fig2}, which has been used as an estimate in Ref.~\cite{achapion}, among other options.

Using the isospin extra factor $\sqrt{2}$ for $D^0\to D^{*+}\pi^-$ relative to $D^0\to D^{*0}\pi^0$, the amplitude of Fig. \ref{fig:fig2} is given by
\begin{equation}
 t^\mathrm{tree}_{D^0\bar{D}^0\to \pi^+\pi^-}=\frac{\tilde{g}^2}{q^2-M^2_{D^*}}(p_1+p_3)\cdot (p_2+p_4)\ ,\label{eq:ttree}
\end{equation}
which is about $45$ times bigger than the result from Eq. (\ref{eq:tx1}). We should also note that the estimate for the $D^{0}\bar{D}^0\to\pi^+\pi^-$ amplitude \cite{achapion} based on the mechanism of Fig. \ref{fig:fig2} is about a factor of five smaller than that from Eq. (\ref{eq:ttree}), and $9$ bigger than the result from Eq. (\ref{eq:tx1}).  This short discussion clearly indicates that, although we think that the amplitude of Eq. (\ref{eq:tx1}) is realistic, we must accept some uncertainties in the strength of the predicted cross sections. But we should also emphasize that the position and shape of the predicted peaks does not depend on this amplitude. In the evaluations done here for the diagrams of Fig. \ref{fig:fig1} we use the amplitude of Eq.~ (\ref{eq:tx1}).
\subsection{The $X\to \pi^0\pi^+\pi^-$ Triangle mechanism}\label{sec:tri}
\begin{figure*}
 \centering
 \includegraphics[scale=0.7]{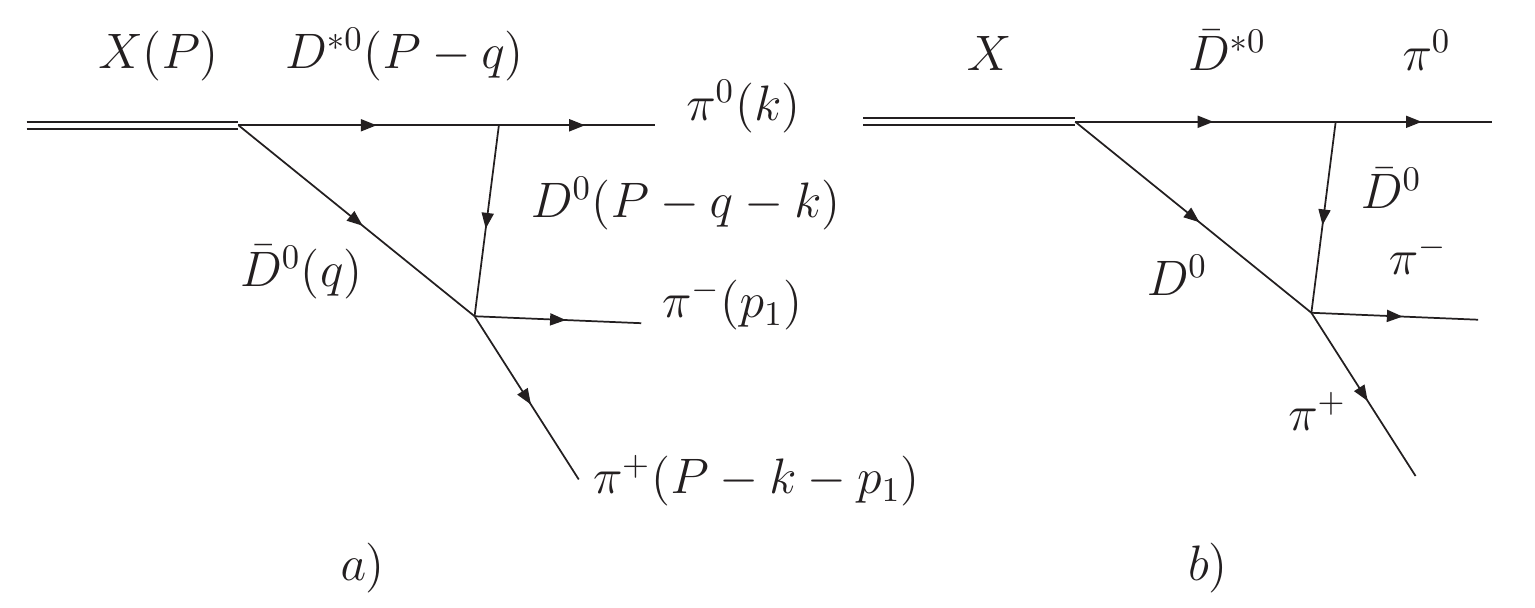}
 \caption{Triangle mechanisms for the $X\to \pi^0\pi^+\pi^-$ process. In diagram a) we show in parenthesis the momenta of the particles.}
 \label{fig:fig3}
\end{figure*}

Prior to the evaluation of the diagrams of Fig. \ref{fig:fig1} we shall evaluate the width of the $X$ going through the two mechanisms depicted in Fig. \ref{fig:fig3}, both of which give identical contribution to the $X\to \pi^0\pi^+\pi^-$ amplitude.

In order to evaluate the amplitude of Fig. \ref{fig:fig3} we make use of the fact that the $D,D^*$ in the intermediate states are placed on shell in the TS. This, and the fact that the particles are heavy, make it unnecessary to take into account the negative energy part of the $D,D^*$ propagators. Hence, 
\begin{eqnarray}
 &&D(q)=\frac{1}{q^2-m^2+i\epsilon}\longrightarrow \frac{1}{2\omega}\frac{1}{q^0-\omega+i\epsilon}\ ,\label{eq:d}
\end{eqnarray}
and only the positive energy part is taken. In Eq. (\ref{eq:d}), $\omega\equiv \omega(q)=\sqrt{\vec{q}\,^2+m^2}$. Since both diagrams in Fig.~\ref{fig:fig3} give rise to the same amplitude, the total amplitude is written by,
\begin{eqnarray}
 &&t=-2ig_X\tilde{g}\int \frac{d^4q}{(2\pi)^4}\frac{1}{2\sqrt{2}}\vec{\epsilon}_X\cdot\vec{\epsilon}_{D^*}\nonumber\\&&\times \vec{\epsilon}_{D^*}\cdot(2\vec{k}+\vec{q})\,t_{D^0\bar{D}^0,\pi^+\pi^-}\frac{1}{2\omega^*}\frac{1}{2\omega_1}\frac{1}{2\omega_2}\nonumber\\&&\times \frac{1}{P^0-q^0-\omega^*+i\epsilon} \frac{1}{q^0-\omega_1+i\epsilon}\nonumber\\&&\times\frac{1}{P^0-q^0-k^0-\omega_2+i\epsilon}\ ,
\end{eqnarray}
where $\omega^*\equiv\omega^*(q)=\sqrt{m^2_{D^*}+\vec{q}\,^2}$, $\omega_1\equiv\omega_1(q)=\sqrt{m_D^2+q^2}$, $\omega_2\equiv\omega_2(\vec{q}+\vec{k})=\sqrt{m^2_D+(\vec{q}+\vec{k})^2}$. Summing over the $D^{*0}$ polarizations and performing the $q^0$ integration analytically using Cauchy's theorem, one obtains
\begin{eqnarray}
 t=-\frac{1}{\sqrt{2}}g_X\tilde{g}\,t_{D^0\bar{D}^0,\pi^+\pi^-}\vec{\epsilon}_X\cdot \vec{k}\,t_T\ ,\label{eq:ttt}
\end{eqnarray}
where $t_T$ is given by
\begin{eqnarray}
 &&t_T=\int\frac{d^3q}{(2\pi)^3}\frac{1}{2\omega^*}\frac{1}{2\omega_1}\frac{1}{2\omega_2}(2+\frac{\vec{q}\cdot\vec{k}}{\vec{k}\,^2})\theta(q_\mathrm{max}-\vert\vec{q}\vert^*)\nonumber\\&&\times \frac{1}{P^0-\omega_1-\omega^*+i\frac{\Gamma_{D^*}}{2}}\frac{1}{P^0-k^0-\omega_1-\omega_2+i\epsilon} \ ,\label{eq:tt}
\end{eqnarray}
where we have added the cut off $\theta(q_\mathrm{max}-\vert\vec{q}\vert^*)$, which comes associated with the approach of Ref.~\cite{danijuan}, and $q^*$ being the momentum of the $\bar{D}^0$ in Fig. \ref{fig:fig3} a) in the $\pi^+\pi^-$ rest frame. To arrive to Eq. \ref{eq:tt}, we have replaced $\int d^3q q_i$ by $\int d^3q\frac{\vec{q}\cdot\vec{k}}{\vec{k}\,^2}k_i$, since the the resulting integration is a vector and the only vector non-integrated is $\vec{k}$. In addition, we have considered the width of the $D^*$ by means of the substitution $\omega^*\longrightarrow \omega^*-i\frac{\Gamma_{D^*}}{2}$.

We take $q_\mathrm{max}=800~\mathrm{MeV}$, and since the momentum $q$ appearing in the TS is of the order of $10~\mathrm{MeV}/c$, we can take $\theta(q_\mathrm{max}-\vert\vec{q}\vert^*)=\theta(q_\mathrm{max}-\vert\vec{q}\vert)$ for practical purposes, since momenta around $q_\mathrm{max}^*$ give negligible contribution to the TS. This allows us to do analytically the $\mathrm{cos}\,\theta$ integration of Eq. (\ref{eq:tt}), and we find, 
\begin{eqnarray}
 &&t_T=\int^{q_\mathrm{max}}_0\frac{q^2dq}{32\pi^2}\frac{1}{\omega^*}\frac{1}{\omega_1}\frac{1}{k\,q}\frac{1}{P^0-\omega_1-\omega^*+i\frac{\Gamma_{D^{*0}}}{2}}\nonumber\\&&\times \left\lbrace(2-\frac{m^2_{D^0}+k^2+q^2-b^2}{2k^2})\mathrm{ln}\frac{b-\omega_2^{-}+i\epsilon}{b-\omega_2^++i\epsilon}\right.\nonumber\\&&
\left.+\frac{1}{2k^2}\left[-\frac{1}{2}(b-\omega_2^+)^2+\frac{1}{2}(b-\omega_2^-)^2+2b(\omega_2^--\omega_2^+)\right]\right\rbrace\ \nonumber\\\label{eq:tt2}\end{eqnarray}
where
\begin{eqnarray}
 &&b\equiv P^0-k^0-\omega_1;\,P^0\equiv M_X;\, q\equiv |\vec{q}|\ ,\nonumber\\&&k^0=\frac{M^2_X+m^2_{\pi^0}-M^2_{\pi^+\pi^-}}{2M_X}\ ,\nonumber\\&&k\equiv|\vec{k}|=\frac{\lambda^{1/2}(M^2_X,m^2_{\pi^0},M^2_{\pi^+\pi^-})}{2M_X}\ ,\nonumber\\&&\omega_2^+=\sqrt{m^2_{D^0}+k^2+q^2+2k\,q}\ ,\nonumber\\&&\omega_2^-=\sqrt{m^2_{D^0}+k^2+q^2-2k\,q}\ .
\end{eqnarray}
The TS appears technically when the denominator in Eq. (\ref{eq:tt2}) becomes zero, $P^0-\omega_1-\omega^*$ (ignoring the $\Gamma_{D^{*0}}$ width), and $b-\omega_2^-$ becomes zero. The presence of the $D^{*0}$ width renders the contribution finite, but given the small value of $\Gamma_{D^{*0}}$, $t_T$ gives rise to very sharp peaks.

The differential mass distribution is given by
\begin{eqnarray}
\frac{d\Gamma}{dM_{\pi^+\pi^-}}=\frac{1}{(2\pi)^3}k\,\tilde{p}_{\pi^+}\bar{\sum}\sum \vert t\vert^2\frac{1}{4M_X^2}\ ,\label{eq:dgam}
\end{eqnarray}
where $\tilde{p}_{\pi^+}=\frac{\lambda^{1/2}(M^2_{\pi^+\pi^-},m^2_{\pi^+},m^2_{\pi^-})}{2M_{\pi^+\pi^-}}$, and,
\begin{eqnarray}
 \bar{\sum}\sum \vert t\vert^2=\frac{1}{6}\tilde{g}^2\vert g_X\,t_{D^0\bar{D}^0,\pi^+\pi^-}\vert^2\vec{k}\,^2\vert t_T\vert^2\ ,
\end{eqnarray}
and $t_T$ is given by Eq. (\ref{eq:tt2}).
\subsection{The $X\to D^{*0}D\to D^0\bar{D}^0\pi^0$ width}\label{sec:xwidth}
\begin{figure*}
 \centering
 \includegraphics[scale=0.7]{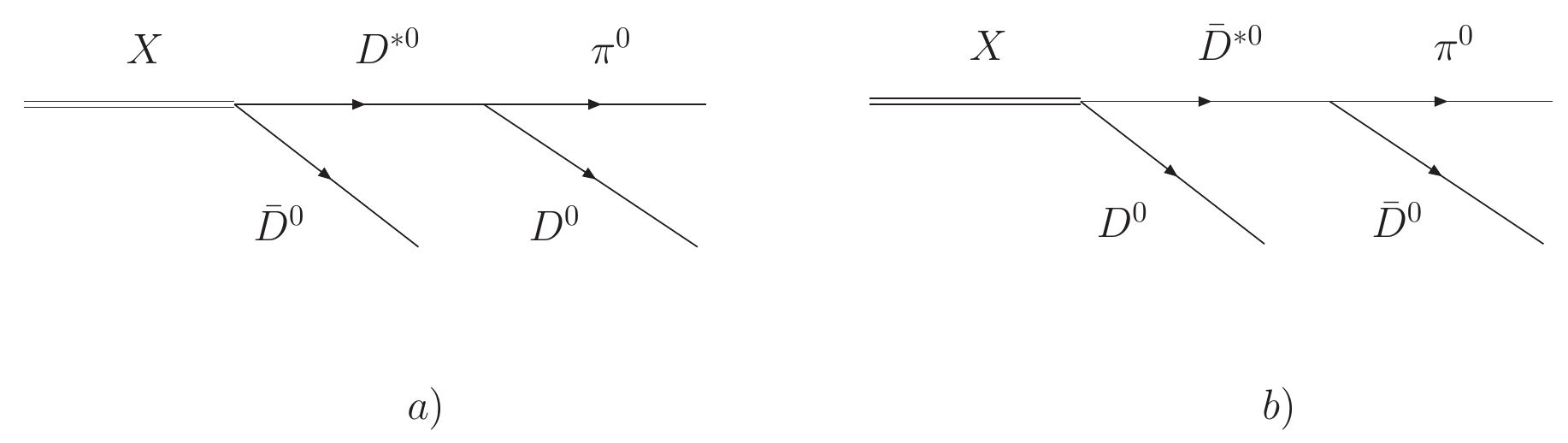}
 \caption{Diagrams for $X\to D^{*0}\bar{D}^0$, $\bar{D}^{*0}D^0$, followed by $D^{*0}(\bar{D}^{*0})$ decay to $\pi D^0(\bar{D}^{*0})$.}
 \label{fig:fig4}
\end{figure*}

The results of Eq. (\ref{eq:dgam}) should be compared with the decay width of the $X$  to the $D^0\bar{D}^{*0}\pi^0$ channel with the given $X,D^{*0}\bar{D}^0$ coupling. The mechanisms for this decay proceed as shown in the diagrams of Fig. \ref{fig:fig4}.

The amplitude for the diagram of Fig. \ref{fig:fig4} a) is given by
\begin{eqnarray}
 &&t_a=-\frac{g_X\tilde{g}}{2\sqrt{2}}\vec{\epsilon}_X\vec{\epsilon}_{D^*}\frac{\vec{\epsilon}_{D^*}\cdot(\vec{p}_\pi-\vec{p}_D)}{M^2_{\pi^0D^0}-M^2_{D^{*0}}+iM_{D^{*0}}\Gamma_{D^{*0}}}\ .\nonumber\\
\end{eqnarray}
Taking into account both diagrams, Fig.~\ref{fig:fig4} a) and b), after summing over the $D^{*0}$, $\bar{D}^{*0}$ polarizations, we obtain,
\begin{eqnarray}
 &&\tilde{t}=-\frac{g_X\tilde{g}}{2\sqrt{2}}\left\lbrace \frac{\vec{\epsilon}_X\cdot (\vec{p}_\pi-\vec{p}_D)}{M^2_{\pi^0D^0}-M^2_{D^{*0}}+iM_{D^{*0}}\Gamma_{D^{*0}}}\right.\nonumber\\&&+\left.\frac{\vec{\epsilon}_X\cdot(\vec{p}_\pi-\vec{p}_{\bar{D}})}{M^2_{\pi^0\bar{D}^0}-M^2_{\bar{D}^{*0}}+iM_{\bar{D}^{*0}}\Gamma_{D^{*0}}}\right\rbrace\ .
\end{eqnarray}
Given the small $D^*$ momenta, the $\pi^0D^0$ or $\pi^0\bar{D}^0$ move freely in any direction and there is no appreciable interference between the two mechanisms, such that we can write,
\begin{eqnarray}
 \frac{d\Gamma}{dM_{\pi^0D^0}}=\frac{1}{(2\pi)^3}p_{\bar{D}^0}\tilde{p}_{\pi^0}\bar{\sum}\sum \vert \tilde{t}\vert^2\frac{1}{4 M^2_X}\ ,
\end{eqnarray}
being 
\begin{eqnarray}
 &&p_{\bar{D}^0}=\frac{\lambda^{1/2}(M^2_X,M^2_{\bar{D}^0},M^2_{\pi^0D}}{2M_X}\ ,\nonumber\\&&\tilde{p}_{\pi^0}=\frac{\lambda^{1/2}(M^2_{\pi^0 D^0},m^2_{\pi^0},m^2_{D^0})}{2M_{\pi^0 D^0}}\ , \nonumber
\end{eqnarray}
 and 
\begin{eqnarray}
 &&\bar{\sum}\sum \vert \tilde{t}\vert^2=\frac{1}{9}\tilde{p}\,^2_{\pi^0}\left\vert \frac{g_X\tilde{g}}{M^2_{\pi^0D^0}-M^2_{D^{*0}}+i M_{D^{*0}}\Gamma_{D^{*0}}}\right\vert^2.\nonumber\\
\end{eqnarray}
\subsection{Amplitude for $B^-\to K^-X$; $X\to\pi^0\pi^+\pi^-$}
By taking the result for the evaluation of the $X\to\pi^0\pi^+\pi^-$ amplitude done in sec. \ref{sec:tri}, in particular Eq. (\ref{eq:ttt}), next we write the amplitude for the first two diagrams of Fig. \ref{fig:fig1},
\begin{eqnarray}
 &&t'=-\frac{1}{\sqrt{2}}{\cal C} g_X\tilde{g}\frac{\vec{\epsilon}_X\cdot\vec{p}_K\,t_{D^0\bar{D}^0,\pi^+\pi^-}\vec{\epsilon}_X\cdot \vec{k}\, t_T}{M^2_{\pi^0\pi^+\pi^-}-M^2_X+iM_X\Gamma_X}\nonumber\\
\end{eqnarray}
where $p_K\equiv|\vec{p}_K|=\frac{\lambda^{1/2}(M^2_B,m_k^2,M^2_{\pi^0,\pi^+\pi^-}}{2M_B}$, and $t_T$ given by Eq. (\ref{eq:tt2}). After summing over the $X$ polarizations,
\begin{eqnarray}
 t'=-\frac{1}{\sqrt{2}}{\cal C}g_X\tilde{g}\frac{\vec{p}_K\cdot \vec{k}\,t_{D^0\bar{D}^0,\pi^+\pi^-} t_T}{M^2_{\pi^0\pi^+\pi^-}-M^2_X+i M_X\Gamma_X}\ .\nonumber\\
\end{eqnarray}
In $|t'|^2$ we consider that the angle average of $\vert\vec{p}_K\cdot \vec{k}\vert^2$ is $\frac{1}{3}|\vec{p}_K|^2\vec{k}\,^2$, and then, one finds
\begin{eqnarray}
 &&\bar{\sum}\sum \vert t'\vert^2=\frac{1}{6}{\cal C}^2 
 \frac{\tilde{g}^2\vert g_X\,t_{D^0\bar{D}^0,\pi^+\pi^-}\vert^2\vec{p}\,^2_K\vec{k}^2\vert t_T\vert^2}{\left\vert M^2_{\pi^0\pi^+\pi^-}-M^2_X+i M_X\Gamma_X\right\vert^2}\nonumber\\
\end{eqnarray}
from where we obtain the double differential mass distribution \cite{pavaosakai},
\begin{eqnarray}
 &&\frac{d^2 \Gamma_{B^-}}{dM_{\pi^0\pi^+\pi^-}dM_{\pi^+\pi^-}}=\frac{p_K p_{\pi^0}\tilde{p}_{\pi^+}}{128\pi^5M^2_{B^-}}\bar{\sum}\sum\vert t'\vert^2,\nonumber\\\label{eq:dgampi}
\end{eqnarray}
where 
\begin{eqnarray}
 &&p_{\pi^0}=\frac{\lambda^{1/2}(M^2_{\pi^0\pi^+\pi^-},m^2_{\pi^-},M^2_{\pi^+\pi^-})}{2M_{\pi^0\pi^+\pi^-}}\ ,\nonumber\\
 &&\tilde{p}_{\pi^+}=\frac{\lambda^{1/2}(M^2_{\pi^+\pi^-},m^2_{\pi^+},m^2_{\pi^-})}{2M_{\pi^+\pi^-}}\ .
\end{eqnarray}
In practice, we perform the integral of Eq. (\ref{eq:dgampi}) over $M_{\pi^0\pi^+\pi^-}$, which results in a convolution of the decay width of the $X$ for the process $X\to \pi^0\pi^+\pi^-$ with the spectral function (mass distribution) of the $X$. The resulting $d\Gamma_{B^-}/dM_{\pi^+\pi^-}$ will show the TS of the mechanism disscussed.
\section{Results}
\subsection{The $X\to D^0\bar{D}^0\pi^0$ width}
\begin{figure}
 \centering
 \includegraphics[scale=0.45]{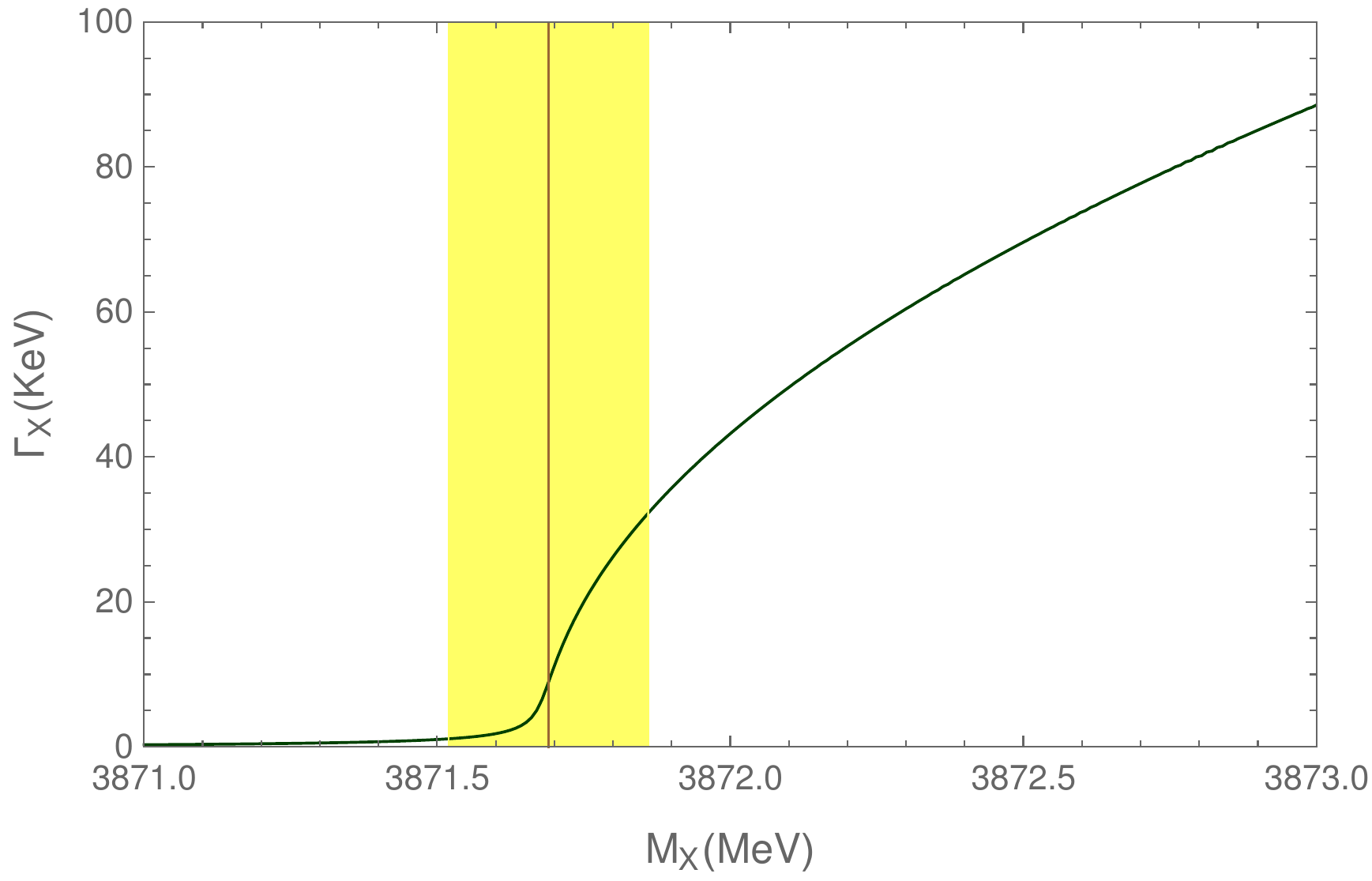}
 \caption{$\Gamma_X$ for $X\to \pi^0D^0\bar{D}^0$ as a function of $M_X$ calculated with $g_X=2~\mathrm{GeV}$. The vertical line and yellow error band represent the X(3872) mass and error according to Eq. (\ref{eq:xmass}).}
 \label{fig:xwidth}
\end{figure}

In Fig. \ref{fig:xwidth} we show the results for $X\to D^0\bar{D}^0\pi^0$ width evaluated in subsec. \ref{sec:xwidth} as a function of the $X$ mass. We can see that around the mass of the $X$ given by Eq. (\ref{eq:xmass}), the width is of the order of $30~\mathrm{KeV}$. This is one source of the width, but according to \cite{guo,guo1}, the total $X$ width cannot be larger than $100~\mathrm{KeV}$. We will perform calculations of Eq. (\ref{eq:dgampi}) adding $50~\mathrm{KeV}$ or $100~\mathrm{KeV}$ to the width of Fig. \ref{fig:xwidth}. In Ref.~\cite{achapion} evaluations are done with $\Gamma_\mathrm{non}$ values of $50~\mathrm{KeV}$ to $200~\mathrm{KeV}$ (the additional $X$ width to the $D^*\bar{D}-cc$ channels) which would be too large according to \cite{guo,guo1} and lead to drastic reductions of the $X\to \pi^0\pi^+\pi^-$ width.

\subsection{The $X\to \pi^0\pi^+\pi^-$ width}
\begin{figure*}
 \centering
 \includegraphics[scale=0.45]{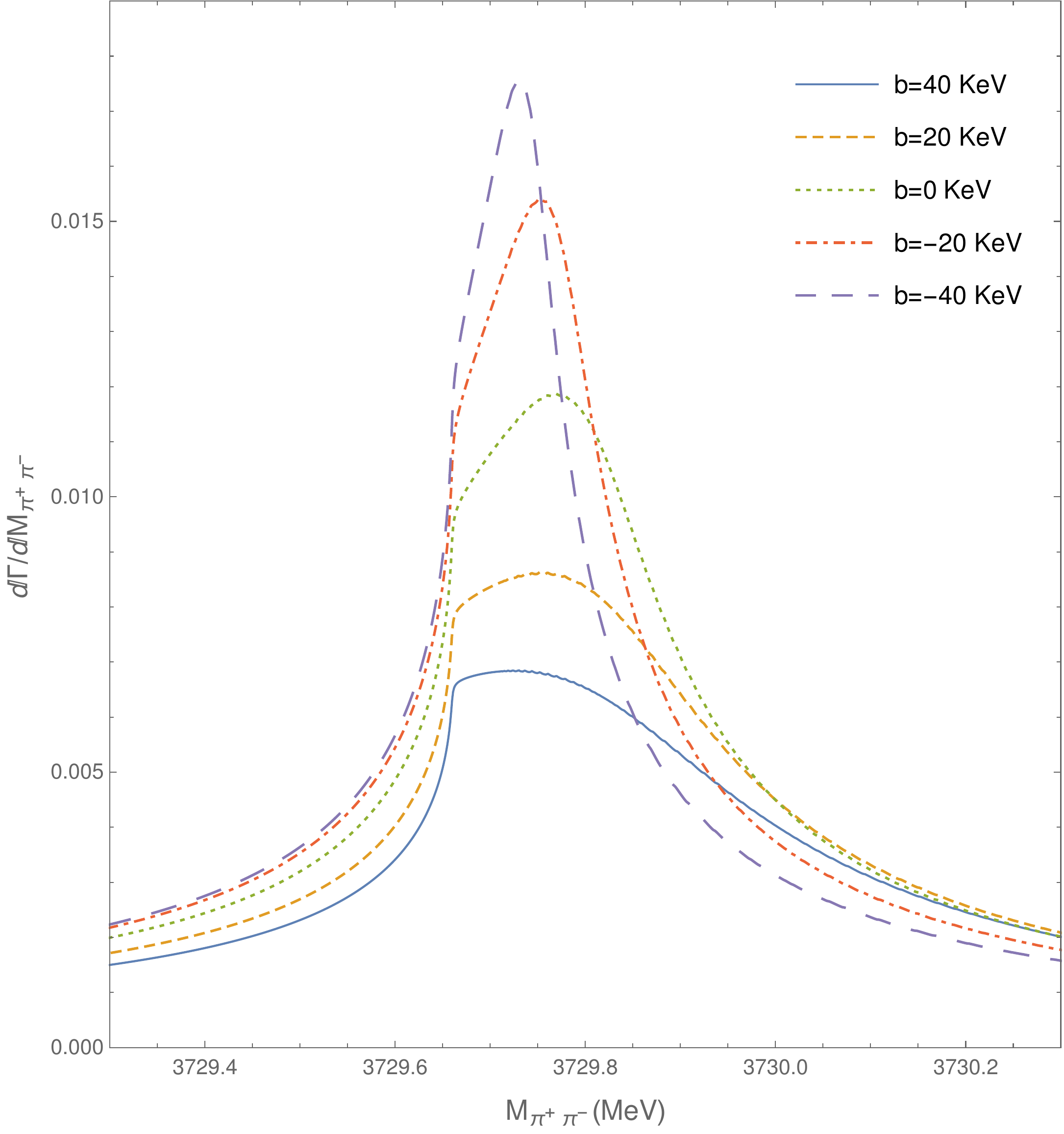}
 \caption{$d\Gamma/d M_{\pi^+\pi^-}$ for different values of $b=m_{D^0}+m_{D^{*0}}-m_X$, where the threshold $m_{D^0}+m_{D^{*0}}=3871.68~\mathrm{MeV}$ \cite{pdg}.}
 \label{fig:dgama}
\end{figure*}

In Fig. \ref{fig:dgama} we show $d\Gamma_X/dM_{\pi^+\pi^-}$ for different values of the $X$ mass. We can see that we obtain peaks for all cases and that changes in $10^{-2}~\mathrm{MeV}$ in the mass of the $X$ change the peak positions of the TS in a similar amount. We can see that in the case of bound or unbound state even the shapes are different. This situation is similar to the one observed in Ref.~\cite{guo} for $D^{*0}\bar{D}^{*0}\to \gamma X$. However we anticipate that in a real reaction, like $B^-\to K^-X\to K^-\pi^0\pi^+\pi^-$, the mass of the $X$
is folded with its spectral function due to its finite width. So, we must see what happens in each particular reaction, and we address this point in the next subsection.

\subsection{The $B^-\to K^-X\to K^-\pi^0\pi^+\pi^-$ mass distributions}
\begin{figure*}
 \centering
 \includegraphics[scale=0.45]{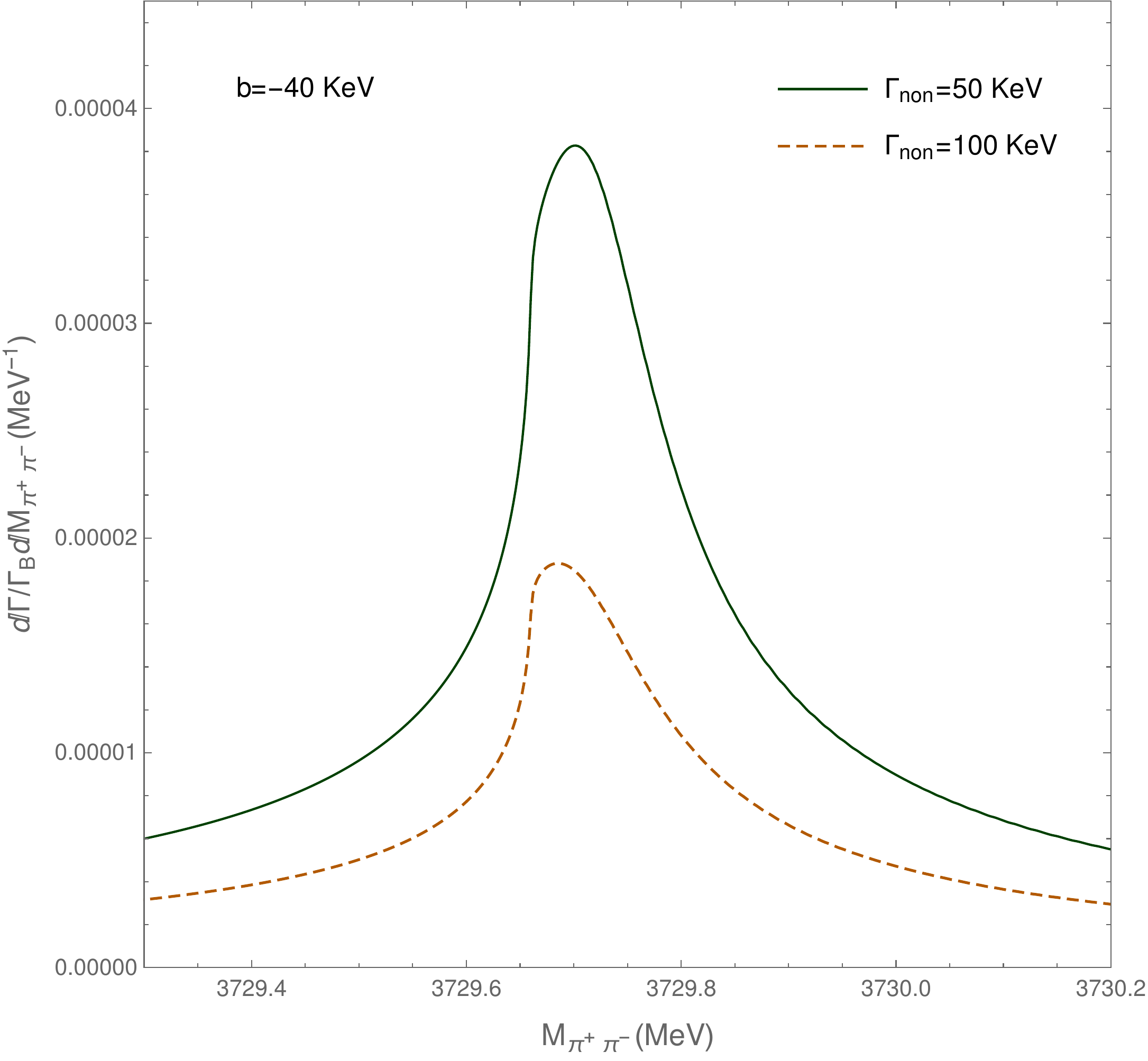}
 \caption{$d\Gamma/(\Gamma_{B^-}d M_{\pi^+\pi^-})$ for $M_X=3871.72$ ($b=-40~\mathrm{KeV}$) and different values of $\Gamma_{\mathrm{non}}$ ($\Gamma_X=\Gamma_X'+\Gamma_\mathrm{non}$.)}
 \label{fig:dgambf}
\end{figure*}

In Fig. \ref{fig:dgambf} we show $d\Gamma_{B^-\to K^-\pi^0\pi^+\pi^-}/dM_{\pi^+\pi^-}$ obtained integrating Eq. (\ref{eq:dgampi}) over $M_{\pi^0\pi^+\pi^-}$ for different values of the added width $50~\mathrm{KeV}$, $100~\mathrm{KeV}$. We can see that an additional $X$ width makes the strength of the distribution smaller but the peak position appears at the same place.

\begin{figure*}
 \centering
 \includegraphics[scale=0.45]{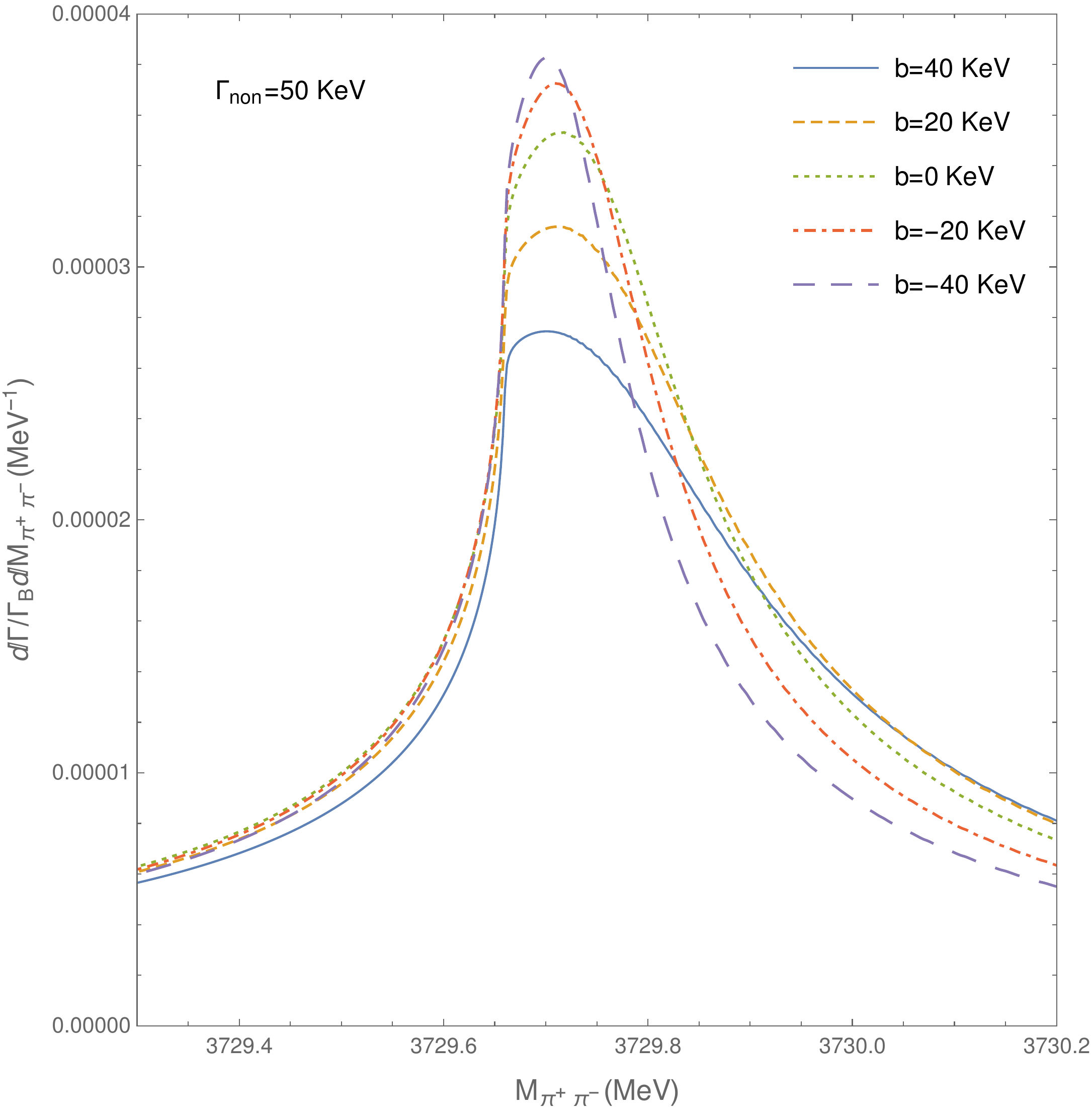}
 \caption{$d\Gamma/(\Gamma_{B^-}d M_{\pi^+\pi^-})$ for $\Gamma_\mathrm{non}=50~\mathrm{KeV}$ and different values of $b=m_{D^0}+m_{D^{*0}}-m_X$.}
 \label{fig:dgamb50}
\end{figure*}

\begin{figure*}
 \centering
 \includegraphics[scale=0.45]{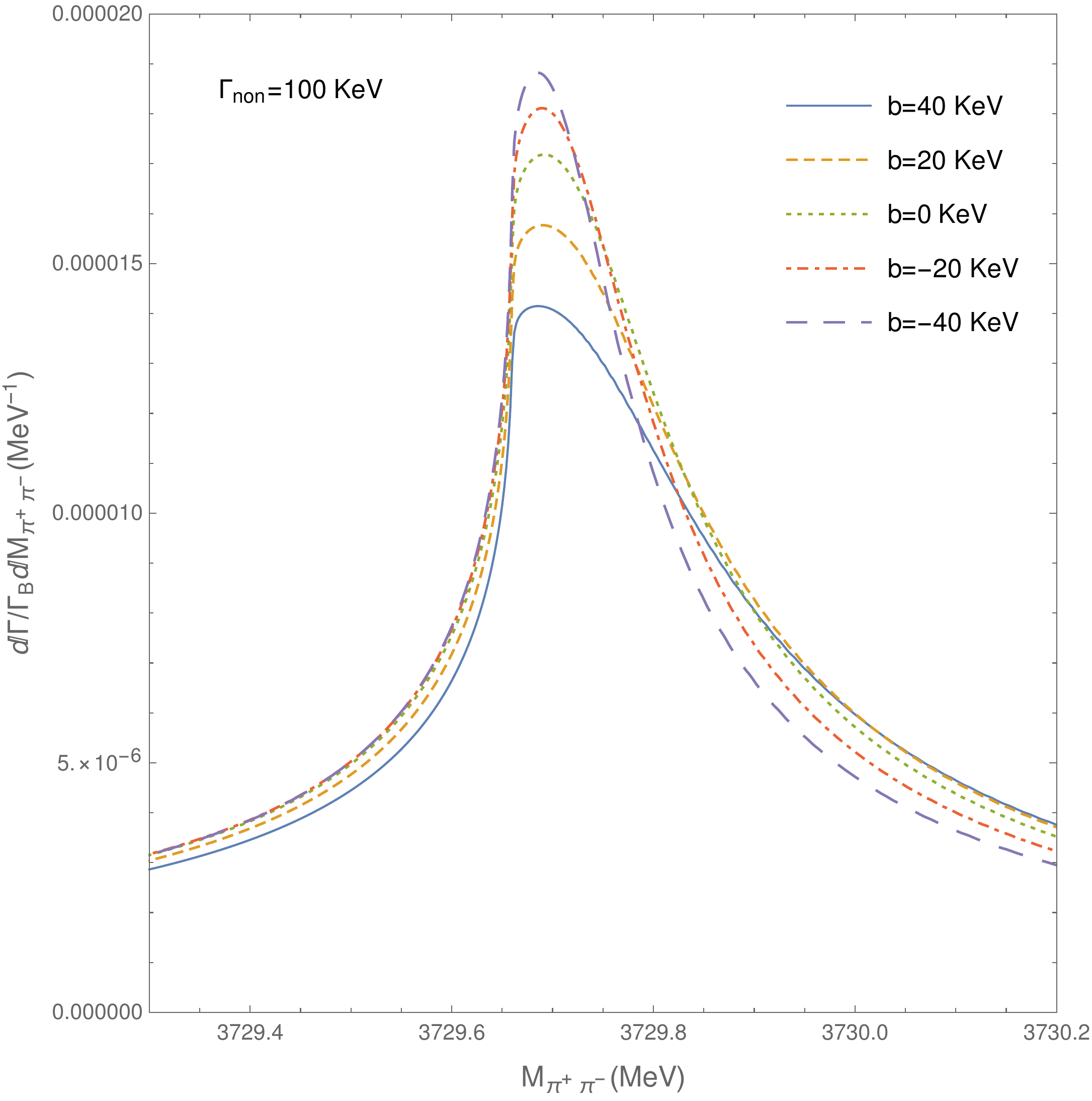}
 \caption{$d\Gamma/(\Gamma_{B^-}d M_{\pi^+\pi^-})$ for $\Gamma_\mathrm{non}=100~\mathrm{KeV}$ and different values of $b=m_{D^0}+m_{D^{*0}}-m_X$.}
 \label{fig:dgamb100}
\end{figure*}

\begin{figure*}
 \centering
 \includegraphics[scale=0.7]{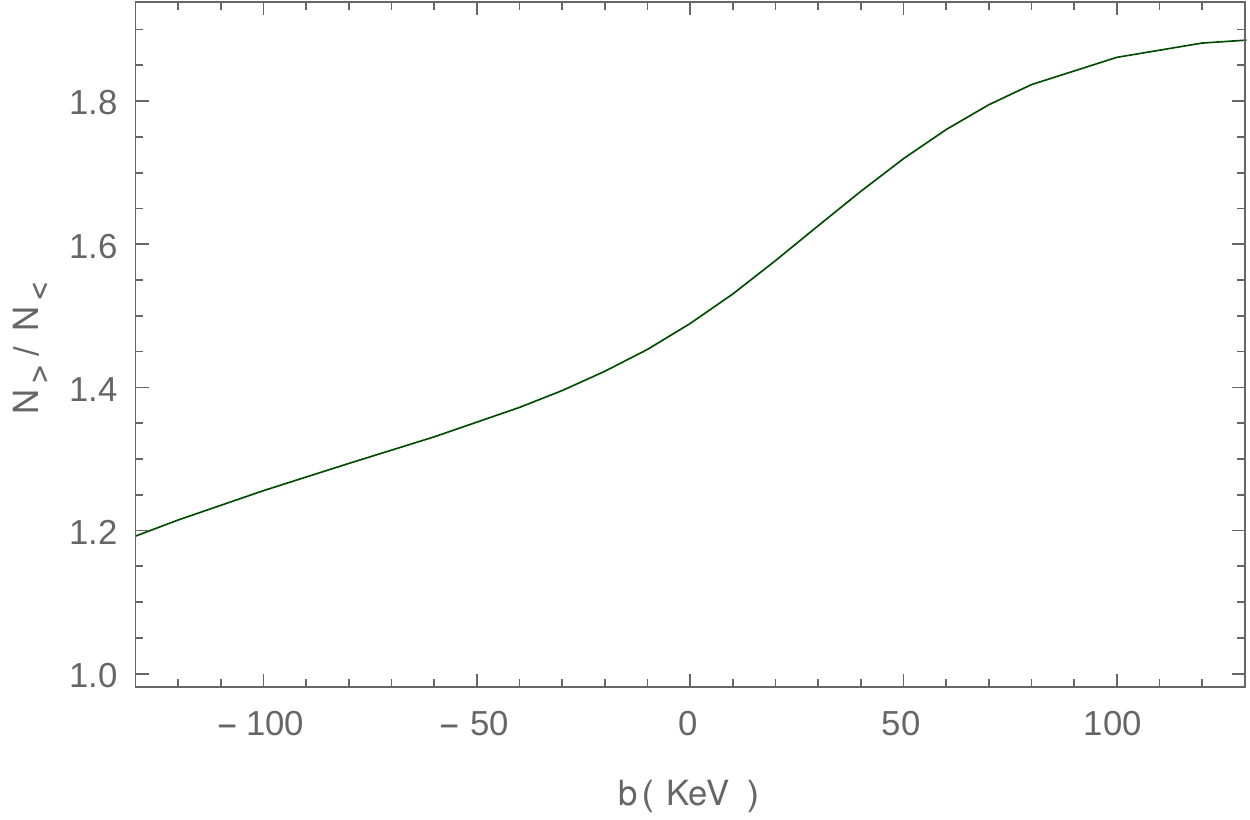}
 \caption{Asymmetry $N_>/N_<$ as a function of the binding energy of the X(3872) for $\Gamma_{\mathrm{non}}=100$ KeV.}
 \label{fig:asi}
\end{figure*}

In Figs. \ref{fig:dgamb50} and \ref{fig:dgamb100} we show the mass distribution for $\Gamma_\mathrm{non}=50$ and $100$~KeV, respectively and different $X$ masses. It can be seen that the shape of the peak is quite sensitive to the binding energy $b$ of the $X$, and the shape is similar in both figures. We follow the same idea as in Ref. \cite{Sakai} and evaluate the asymmetry as
\begin{equation}
 \frac{N_>}{N_<}\equiv\frac{\int^{\hat{M}+\delta}_{\hat{M}}dM_{\pi^+\pi^⁻}\frac{d\Gamma}{\Gamma_{B^-}dM_{\pi^+\pi^-}}}{\int^{\hat{M}}_{\hat{M}-\delta}dM_{\pi^+\pi^⁻}\frac{d\Gamma}{\Gamma_{B^-}dM_\mathrm{\pi^+\pi^-}}}\ ,
\label{eq:asi}\end{equation}
where $\hat{M}$ is the value of the invariant mass of the two pions where the differencial distribution inside the integral takes its maximum and $\delta=0.5~\mathrm{MeV}$ to cover the full strength of the peak. This is depicted in Fig.~\ref{fig:asi} for $\Gamma_\mathrm{non}=100$ KeV. As can be seen,  the shape of the distribution becomes quite asymmetric for a bound state, while it is closer to being symmetric if the $X(3872)$ is a resonance. The difference of the asymmetry of Eq. (\ref{eq:asi}) between $1.25$ and $1.85$ for $b=-100~\mathrm{ KeV}$ and $100~\mathrm{ KeV}$ is quite large, and even $50~\mathrm{ KeV}$ difference in the binding should lead to  observable effects. 


The interesting thing about this proposal is that in order to determine the $X$ mass one does not have to measure the $\pi^0\pi^+\pi^-$ invariant mass and hence one does not have to measure the $\pi^0$, although its detection would serve to reduce background. Similarly, one would not even need to measure the $K^-$, although again detecting it will reduce the background.

So, let us say that we detect the $K^-$ and the $\pi^0$ (no precision is needed) and we measure only the $\pi^+\pi^-$ invariant mass with precision. Determining the peak of the TS tells us which is the $X$ mass. Given that $\pi^+\pi^-$ can be measured with high precision one could anticipate that the $X$ mass could be determined with a better precision than the present one. The rates for production are not too small. Integrating the peaks of Fig. \ref{fig:dgamb100} over $M_\mathrm{\pi^+\pi^-}$ we obtain a branching fraction of $4\times 10^{-6}$, about a factor $50$ smaller than the ${\cal BR}$ for $B^-\to K^-X$.
\\
\section{Conclusions}
We have studied the width of the $X(3872)$ due to a triangle mechanism that generates a triangle singularity, with peculiar features highly sensitive to the $X(3872)$ mass. The mechanism is given by the decay of the $X$ into $D^{*0} \bar{D}^0 -cc$, with the $D^{*0} (\bar{D}^{*0})$ decaying to $D^0 \pi^0 (\bar{D}^0 \pi^0)$. In a third step the $D^0 \bar{D}^0$ interact producing a pair of pions, $\pi^+ \pi^-$. We find that this mechanism gives rise to a very narrow peak in the invariant mass of the  final $\pi^+ \pi^-$. The asymmetry of the peak is very sensitive to the precise value of the $X$ mass, such that its experimental determination indirectly gives the $X$ mass. We take advantage of it and define the asymmetry of the distribution counting events to the right and the left of the peak, and the ratio of these two magnitudes is very sensitive to the $X$ mass. Since this involves integrated rates and has more statistics, this magnitude could turn out to be the best suited to determine the $X$ mass. 

  The interesting thing is that, given the relationship between the $X$ mass and the peak in the  $\pi^+ \pi^-$ invariant mass, one only has to measure the $\pi^+$ and $\pi^-$ with high precision, and these particles can indeed be measured very precisely. Formally the $K^-$ and the $\pi^0$ do not have to be measured because the relationship of the $X$ mass to $M_{\pi^+ \pi^-}$ does not depend on the energy of these two particles. Actually, they are defined at the peak of the triangle singularity. In practice the measurement of these two particles is necessary to reduce background, but the precise measurement is unnecessary. It is enough to know that these two particles are produced.

   The rates obtained are relatively large, such that the measurement can be carried in present facilities and certainly will become more amenable in future upgrades. Given the fact that present uncertainties in the $X$ mass do not allow us to know whether the $D^{*0}\bar{D}^0$ component is bound or not, any idea, like the present one, that helps remove this ambiguity should be most welcome. 
   \section*{Acknowledgments}
 This work is partly supported by the Spanish Ministerio de Economia y Competitividad and European FEDER funds under Contracts No. FIS2017-84038-C2-1-P B and No. FIS2017-84038-C2-2-P B, and the project Severo Ochoa of IFIC, SEV-2014-0398, and by the Talento Program of the Community of Madrid, under the project with Ref. 2018-T1/TIC-11167.
\bibliographystyle{plain}

\end{document}